\documentclass[11pt]{article} 
\usepackage{graphicx} 
\usepackage{array}
\usepackage{verbatim}
\usepackage{float}
\usepackage[latin9]{inputenc}
\usepackage{amsthm}
\usepackage{amsmath}
\usepackage{amssymb}
\usepackage{setspace}
\usepackage{esint}
\usepackage{hyperref}

\usepackage{algorithm}
\usepackage[noend]{algpseudocode}

\usepackage{natbib}

\makeatletter
\theoremstyle{plain}
\newtheorem*{thm*}{\protect\theoremname}

\usepackage[T1]{fontenc}
\usepackage{esint}
\usepackage{booktabs}

\usepackage{authblk}

\makeatletter

\theoremstyle{plain}

\theoremstyle{definition}

\usepackage{boxedminipage}

\usepackage{listings}

\usepackage{ifpdf}

\usepackage{longtable}
\usepackage{mathtools}

\usepackage{booktabs}

\date{}

\makeatother

\usepackage{babel}
\providecommand{\definitionname}{Definition}
\providecommand{\theoremname}{Proposition}

\makeatother

\begin{document}

\title{Optimal Designs of Two-Phase Case-Control Studies for General Predictor Effects}

\author[1, 3]{Jingjing Zou}

\author[2]{Lori B.Daniels}

\author[1, 3]{Karen Messer}

\author[4]{Daniel Rabinowitz}

\affil[1]{Herbert Wertheim School of Public Health and Human Longevity Science, University of California, San Diego}

\affil[2]{Division of Cardiovascular Medicine, Department of Medicine \&  Division of Epidemiology, Department of Family Medicine and Public Health, University of California, San Diego}

\affil[3]{Moores Cancer Center, University of California, San Diego}	

\affil[4]{Department of Statistics, Columbia University}	

\date{}

\maketitle

\begin{abstract}
Under two-phase designs, the outcome and several covariates and confounders are measured in the first phase, and a new predictor of interest, which may be costly to collect, can be measured on a subsample in the second phase, without incurring the costs of recruiting subjects. By using the information gathered in the first phase, the second-phase subsample can be selected to enhance the efficiency of testing and estimating the effect of the new predictor on the outcome. Past studies have focused on optimal two-phase sampling schemes for statistical inference on local ($\beta = o(1)$) effects of the predictor of interest. In this study, we propose an extension of the two-phase designs that employs an optimal sampling scheme for estimating predictor effects with pseudo conditional likelihood estimators in case-control studies. This approach is applicable to both local and non-local effects. We demonstrate the effectiveness of the proposed sampling scheme through simulation studies and analysis of data from 170 patients hospitalized for treatment of COVID-19. The results show a significant improvement in the estimation of the parameter of interest.

\end{abstract}

\noindent%

\section{Introduction}

A well-characterized cohort undergoing longitudinal observation can provide
opportunities for cost-efficient ancillary studies.
While the cost of collecting a new predictor may be substantial, the new predictor may
be compared to previously collected outcomes, adjusting for other previously collected covariates
and potential confounders without incurring the costs of recruiting subjects and 
obtaining the previously collected data.  
In some such settings, the previously collected outcome may record the occurrence of a low-probability
event.  
When this is the case, it is natural to consider a two-phase case-control study. 
Information of the outcome, covariates and potential confounders collected in the first phase can be used to select a subsample of subjects from the cases and the controls in the second phase for ascertainment of the new predictor. 
This design can significantly reduce the cost of collecting the new predictor and increase efficiency in statistical inference.
A natural question is: what is the optimal design for selecting case and control subjects in the sub-sample.

% refs
Studies on the design of two-phase studies date back to \cite{WHITE1982}, in which a rare disease and a rare exposure were collected in the first stage and covariates were collected in the second stage based on grouping of the disease and exposure status.
Since then, researchers have proposed various sampling schemes for two-phase studies, depending on the type of outcome and research question of interest.
\cite{Breslow1999} examined a sampling scheme stratifying on both the covariates and a discrete outcome.
\cite{Cao2021} studied two-phase designs with emphasis on measures of predictive accuracy.
\cite{Lin2013, Zhou2014} proposed sampling schemes for two-phase studies with continuous outcomes.
\cite{Gravio} examined the design and analysis of two-phase studies with multivariate longitudinal data.

Among the first to examine optimality for statistical inference in two-phase designs,
\cite{Tao2020} proposed sampling schemes for semi-parametric efficient testing and estimating of local effects ($\beta = o(1)$) of the predictor of interest on different types of outcome.
For local effects, the optimality in design was defined with respect to maximum Fisher information for the regression parameter associated with the new predictor at the null hypothesis, in a regression of the previously collected outcome on the predictor and previously collected covariates and potential confounders.
\cite{Tao2021} further studies a two-wave two-phase design for longitudinal outcomes. 
For studies on two-phase designs for other types of outcomes, see \cite{Tao2020} and the reference therein.

% shrink this paragraph

The choice of optimal two-phase design for estimating general effects of predictors on outcomes, including both local and non-local effects, depends on the estimation method used. Therefore, studies focused on estimating model parameters in two-phase designs are relevant.
Efficient augmented inverse probability weighting estimators for designs were developed by \cite{Robins1995}, but they require strictly positive selection probabilities in the second phase and can be challenging to implement in practice, as solutions to infinite-dimensional integral equations are involved.
Estimators developed for two-phase designs with discrete covariates include
\cite{scott1991, Scott1997, Breslow1997, Lawless1999, Breslow2003, Chatterjee2003, Weaver2005, Dai2009, Cao2020}. 
\cite{Tao2017} proposed semiparametric estimators for two-phase studies with either discrete or continuous covariates and outcome. Their method was based on approximating the conditional density functions of predictor given other covariates using B-spline sieves.
However, this approach can become challenging to implement and computationally burdensome when dealing with a large number of covariates.
\cite{Scott2011, Che2021} studied improvement of efficiency of conditional maximum likelihood estimators.
For estimation methods concentrating on other aspects of two-phase studies, including measurement errors and multi-category outcomes, see 
\cite{Tao2021, Lotspeich2021, Maronge2021} and the references therein.

Estimation methods for predictors missing at random can also be used in two-phase designs when selection probabilities only depend on the outcome and covariates observed in the first phase.
In particular, \cite{Wang1997} and \cite{Wang2002} extended estimators by \cite{Breslow1988} to semiparametric estimators in logistic regressions based on pseudo conditional likelihood (PCL).
These semiparametric estimators do not assume any model for the selection probabilities or the conditional distribution of the predictor given the outcome and covariates/confounders.
With minor modifications, they can handle both continuous and discrete covariates and are straightforward to implement.

In this paper we focus on optimal two-phase case-control designs for estimating the effect of the predictor on the outcome. We propose a novel approach for selecting subjects from the cases and controls that leads to minimal asymptotic errors using the semiparametric PCL estimators in \cite{Wang1997} and \cite{Wang2002}. 
%These estimators balance implementation complexity with statistical inference efficiency.
Our proposed approach extends current methods, including \cite{Tao2020}, to general scenarios where the effect of the predictor can be either local or non-local, and the covariates can be discrete or continuous. 
This is one of the first works on optimal two-phase designs for estimating the effects of predictors, particularly non-local effects. We derive equations for optimal sampling probabilities of subjects given their values of outcome and covariates/confounders collected in the first phase. We further implement an algorithm to estimate the sampling probabilities that achieve asymptotic optimality in the estimation of the predictor effect under any desired sample size constraints.

Additionally, we present both theoretical and empirical comparisons of the optimal sampling schemes proposed in this paper and those in \cite{Tao2020}. We closely examine the assumptions required for each approach and reveal an interesting trade-off between model assumptions and statistical efficiency. We show that increased efficiency can be achieved by assuming a small amount of extra knowledge. Specifically, to use our proposed approach for statistical inference on potentially non-local predictor effects, the conditional first and second moments of the predictor given the covariates, rather than just the conditional variance required by \cite{Tao2020}, must be estimated or known from previous studies. We also investigate the conditions under which the two sampling schemes can be unified with equal optimal selection probabilities.

We assess the effectiveness of our proposed sampling schemes in estimating the parameter of interest and conducting hypothesis testing using various estimation and testing methods. To achieve this, we conduct comprehensive simulation studies using synthetic data and compare our proposed sampling scheme to multiple existing approaches, including the approach proposed in \cite{Tao2020}.
We further demonstrate the proposed approach using data from 170 patients hospitalized for treatment of coronavirus disease 2019 (COVID-19) at the University of California San Diego Health. Results from simulations using both synthetic and real data indicate that our proposed approach can lead to substantial improvements in estimating the effect of the predictor and has practical utility in various scenarios.

The paper is structured as follows. In Section \ref{sect: selection}, we introduce the model and a modified version of the PCL estimators, and derive the optimal sampling schemes. In Section \ref{sect: test and estimate}, we discuss two existing methods for hypothesis testing of the predictor effect with the sub-sample selected using our proposed sampling schemes. In Section \ref{sect: simulation} we conduct comprehensive simulation studies using synthetic data to compare the proposed sampling scheme with other methods when combined with different estimators and testing methods. In Section \ref{sect: data analysis}, we demonstrate and evaluate the proposed method with data from patients hospitalized for treatment of COVID-19. Additional results are provided in the Supplementary Material.

\section{Optimal Selection of Subjects} \label{sect: selection}

\subsection{Model Notation}

Let $X_i$, $Y_i$, and $Z_i$ denote,
respectively, a predictor of interest, a Bernoulli
outcome, and a vector of covariates and potential
confounders associated with the $i^{th}$ of $n$ subjects in a
randomly sampled cohort.
Let $X$, $Y$, and $Z$ denote a generic triplet,
and let $f_Z$, $f_{X|Z}$, and
$f_{Y|X,Z}$ denote, respectively, the marginal density of $Z$, 
the conditional density of $X$ given $Z$, and the conditional probability mass function
of $Y$ given $X$ and $Z$.  
For simplicity, suppose that $f_{Y|X,Z}$ is
defined implicitly by
$\mbox{logit}(f_{Y|X,Z}(y|x,z))=z^T\alpha+x\beta$.
A logistic intercept may be included in the model
by augmenting the covariates $z$ with a constant component.
Let $\pi(z)$ denote the conditional probability that $Y$ is equal to one given that $Z=z$.
Assume that $\pi(z)$ is less than $1/2$ for all $z$.

The $Y_i$ and $Z_i$ are observed in the whole
cohort, but the $X_i$ are observed in only a sub-sample selected by the investigator.   Let $\delta_i$ denote
the indicator that the $i^{th}$ subject
is selected for the sub-sample.
The investigator has access to the $Z_i$ and $Y_i$ when choosing which subjects to include in the subset, but not
the $X_i$.  That is, the $\delta_i$
are conditionally independent of the $X_i$ given
the $Z_i$ and $Y_i$.

Define sampling probabilities $\eta_y(z) = P(\delta_i = 1 \mid Y_i=y, Z_i = z)$ for $y = 0,1$ and
$\mu(z) = P(\delta_i=1 \mid Z_i=z )$.  
Given the observed covariates $Z_i=z$, $\mu(z)$ gives the sampling probability/weight to include the $i$th subject in the sub-sample, while 
$\eta_y(z) $ further specifies the sampling weight to include the $i$th subject in the sub-sample given the observed $Z_i$ and outcome $Y_i=y$, where $y$ is either $1$ or $0$.
Note that the $\eta_y$ and $\mu$ are
assumed the same across subjects; this corresponds
to selection strategies equivariant with respect to re-ordering the indices of the subjects.

%Let $p_y(z) = P(Y_i=y | \delta_i=1, Z_i=z)$ denote the conditional probability of $Y_i$ given $Z_i$ and that the $i$th subject is selected.  
%Note that, under the null hypothesis,
%\begin{equation} \label{eq: mu and pi}
%\mu(z)\;=\;\eta_0(z)(1-\pi(z))+\eta_1(z)\pi(z).
%\end{equation}
%and
%$p_1(z)=1-p_0(z)=\eta_1(z)\pi(z)/\mu(z)$.  

In what follows, we discuss the optimal sampling schemes and the corresponding optimal sampling probabilities $\mu(Z)$ and $(\eta_1(Z), \eta_0(Z))$. 
We focus on the scenario in which the researcher is interested in estimating the effect of the unobserved predictor of interest $X$ on the outcome $Y$.
Here we develop the optimal sampling schemes for two pseudo conditional likelihood (PCL) estimators in \cite{Wang1997, Wang2002}, with some modifications to enable their applications to both continuous and discrete covariates and confounders.
In addition, we provide a detailed discussion on the difference in optimal sampling probabilities 
between the proposed method and the method in \cite{Tao2020} for testing and estimating local effects,
and provide a condition under which the two sampling schemes are unified.

\subsection{The Modified PCL Estimators} \label{sect: PCL estimators}

We provide a brief introduction to the two PCL estimators and explain the modifications we have made to enable their application to a wider range of covariates and confounders beyond those of discrete distributions.
The first estimator, which we refer to as the \textit{PCLvalidate} estimator, only uses data of subjects selected in the ``validation set'', which contains subjects with both $Z$ and $X$ observed after sampling in the second-phase of study. 
The coefficients in the logistic regression are estimated by solving the estimating equations
\begin{equation} \label{eq: PCLval}
	n^{-1/2} \sum_{i=1}^n \delta_i \mathcal{X}_i \Bigg[Y_i - H\Bigg(\mathcal{X}_i \cdot (\alpha_0, \beta, \alpha_Z)^T + \log\bigg(\frac{\eta_1(Z_i)}{\eta_0(Z_i)} \bigg) \Bigg) \Bigg] = 0,
\end{equation}
where $\mathcal{X}_i = (1, X_i, Z^T_i)$, 
$H(u) = (1+ \exp(-u))^{-1}$ is the logistic distribution function, $(\alpha_0, \beta, \alpha_Z)$ are the regression coefficients to be estimated associated with the intercept, $X$ and $Z$, respectively.
$\eta_1(Z_i)$ and $\eta_0(Z_i)$ are the sampling probabilities of selecting a subject with $Y_i=1$ or $Y_i=0$, respectively, given the value of $Z_i$.

The second estimator, which we refer to as the \textit{PCLboth} estimator, uses data from both the ``validation set'' and ``non-validation set'', in which the latter is the subset of subjects with $X$ remaining unobserved after the second phase sampling. 
The coefficients are estimated by solving the estimating equations
\begin{equation*}
	n^{-1/2} \sum_{i=1}^n \Big[ \delta_i \mathcal{X}_i \big\{ Y_i - \hat{H}_{+}({X}_i, Z_i ) \big\} 
	+ (1 - \delta_i) {\mathcal{T}}_i \big\{ Y_i - \hat{H}_{-}(Z_i) \big\} \Big]
	= 0,
\end{equation*}
where ${\mathcal{T}}_i = (1, {R}(Z_i), Z^T_i)$ with ${R}(Z_i)=\log(E[\exp(\beta X_i) | Z_i, Y_i = 0])$. 
The functions ${H}_{+}$ and ${H}_{-}$ are defined by
\begin{equation*}
	{H}_{+}({X}_i, Z_i) = H(\mathcal{X}_i \cdot (\alpha_0, \beta, \alpha_Z)^T + \log(\eta_1(Z_i)/ \eta_0(Z_i)))
\end{equation*}
and  
\begin{equation*}
	{H}_{-}(Z_i) = H({\mathcal{T}}_i \cdot (\alpha_0, 1, \alpha_Z)^T + \log[ (1-\eta_1(Z_i))/(1-\eta_0(Z_i))] ),
\end{equation*}
respectively.

In solving for the PCLboth estimates, $R(Z_i)$ is unknown and needs to be estimated using data in the validation set. \cite{Wang2002} considered the case of discrete $Z$ and estimated $R(Z_i)$ based on the empirical conditional distribution of $X$ given $Z_i$ and $Y_i=0$.
This approach cannot be applied directly to continuous $Z$.
Here we propose a modified version of PCLboth so that the method is applicable to both discrete and continuous $Z$.
In estimating ${R}(Z_i)$, instead of integrating over the empirical conditional distribution of $X$ given $Z$ and $Y$ in the validation set as in \cite{Wang2002},
a parametric linear regression model can be fitted for continuous $X$ and a logistic regression can be fitted for binary $X$ on $Z$ and the indicator function $I(Y = 0)$. 
Since the quantity $R(Z_i) = \log(E[\exp(\beta X_i) | Z_i, Y_i = 0])$ is the logarithm of the moment generating function of the conditional distribution of $X$ given $(Z, I(Y = 0))$, assuming a model for $X$ on $Z$,
we can write $R(Z_i)$ explicitly with the regression parameters and estimate it using the parameter estimates from the validation set. 
For continuous $X$, $R(Z_i)$ can be estimated by, for a fixed value of $\beta$,
\[
\hat{R}(Z_i) = \beta ((Z_i^T, I(Y_i=0)) \cdot \hat{\gamma}) + \beta^2 \hat{\sigma}_X^2/2,
\] 
where $\hat{\gamma}$ and $\hat{\sigma}_X^2$ are the estimated regression coefficients and standard error in the linear regression.
For binary $X$, $R(Z_i)$ can be estimated by
\[
\hat{R}(Z_i) = \log \big(1-\hat{p}(Z_i) + \hat{p}(Z_i)\cdot \exp(\beta)\big),
\]
where $\hat{p}(Z_i)$ is the predicted probability of $X_i=1$ given $Z_i$ and $Y_i = 0$ in the logistic regression.

\subsection{Optimal Subample Selection for Estimation of Effect of $X$} \label{sect: optimal_PCL}

Both PCLvalidate and PCLboth estimators have been shown to be consistent and follow asymptotic normal distributions (\cite{Wang1997, Wang2002}).
Optimal sampling schemes are defined as those that result in the smallest asymptotic variances of the estimate $\hat{\beta}$ of the coefficient associated with $X$ in the PCL estimators.

For the PCLvalidate estimator, the asymptotic covariance matrix of all the regression coefficient estimates, calculated from the estimating equation \eqref{eq: PCLval}, is (\cite{Wang2002})
\begin{equation} \label{eq: var PCLval}
	\Sigma_{\text{PCLval}}= [E(\delta \mathcal{X}  \mathcal{X}^T H^{(1)}_{+}(X, Z) )]^{-1},
\end{equation}
where $H^{(1)}_{+} = H_{+}(X,Z)(1-H_{+}(X,Z))$.
Given the conditional independence between $\delta$ and $X$ given $Z$, \eqref{eq: var PCLval} can be further written as the inverse matrix of
\begin{equation} \label{eq: cov matrix}
	E \left( \mu(Z)
	\begin{bmatrix}
		\tilde{Z} \tilde{Z}^T E(H^{(1)}_{+} | Z) & \tilde{Z}  E(X H^{(1)}_{+} | Z) \\
		E(X H^{(1)}_{+} | Z) \tilde{Z} ^T & E(X^2 H^{(1)}_{+} | Z)
	\end{bmatrix}
	\right),
\end{equation}
where $\tilde{Z}  = (1, Z^T)^T$.
For convenience, we re-arrange $\mathcal{X}\ = (1, Z^T, X)$ without affecting the result, so that the element associated with the variance of $\hat{\beta}$ is in the bottom right block of the matrix.
Note that  
$H_{+}(X, Z) = H( \mathcal{X} (\alpha_0, \beta, \alpha_Z)^T+ \log(r(Z))) = r(Z)/(r(Z) + \exp(\mathcal{X} (\alpha_0, \beta, \alpha_Z)^T))$ depends on $X$, $Z$ and $r(Z): = \eta_1(Z) / \eta_0(Z)$, which involves the control variables $\eta_1(Z)$ and $\eta_0(Z)$ in the optimization.  
To simplify the optimization, we approximate $H_{+}(X, Z)$ with 
$H_+ := r(Z)/(r(Z) + (1- \pi(Z))/\pi(Z) ) $,
where $\pi(Z) = P(Y=1|Z)$,
and let $H_+^{(1)} = H_+ (1- H_+)$.
Then $H_+^{(1)} $ and $H_+ $ do not depend on $X$ and can be taken out from the conditional expectations. 

Taking the inverse of the matrix \eqref{eq: cov matrix}, the asymptotic variance of $\hat{\beta}$ is given by the reciprocal of 
\begin{multline} \label{eq: var of beta}
	\int \mu(z) H^{(1)}_{+} E(X^2 | z) f(z) dz  \\
	- \int \mu(z) H^{(1)}_{+} \tilde{z}^T  E(X | z) f(z) dz 
	  \left[ \int \mu(z) H^{(1)}_{+} \tilde{z}\tilde{z}^T f(z) dz \right]^{-1}
	 \int \mu(z) H^{(1)}_{+} \tilde{z}  E(X | z) f(z) dz.
\end{multline}
Therefore, to minimize the variance of $\hat{\beta}$, one needs to maximize \eqref{eq: var of beta}, which is
approximately linear with respect to $\mu(z)$ and $H^{(1)}_+$.
Since $\mu(Z) = \eta_1(Z) \pi(Z) + \eta_0(Z) (1-\pi(Z))$, and
$H_{+} = (\eta_1(Z) \pi(Z))/(\eta_1(Z) \pi(Z) + \eta_0(Z) (1-\pi(Z)))$,
$\mu(z) H^{(1)}_{+}$ is maximized, when fixing $\mu(Z)$, by 
\begin{equation} \label{eq: max mu H}
	\mu(Z) \bigg(\frac{1}{4} I(\mu(Z) \in (0, 2\pi(Z)] ) + 
	\bigg(\frac{\pi(Z)}{\mu(Z)} - \frac{\pi^2(Z)}{\mu^2(Z)}\bigg) I(\mu(Z) \in (2\pi(Z),1]) \bigg). 
\end{equation}
Calculating the derivative of \eqref{eq: var of beta} with respect to $\mu(Z)$ and $H^{(1)}_+$, under the constraint $\int f_Z(z)\mu(z)dz = N/n$ of the proportion of subjects selected in the second phase,
we have the Kuhn-Tucker conditions for a fixed Lagrangian multiplier $\lambda$:
\begin{eqnarray} \label{eq: soln of mu2}
	\mu(z)\;=\;0&\mbox{if}
	&{\tilde{\sigma}^2(z)}/{4}< \lambda,\\ \nonumber
	\mu(z) =\;\tilde{\sigma}(z)\pi(z)/ \sqrt{\lambda}
	&\mbox{if}
	&\tilde{\sigma}^2(z)\pi^2(z)\le\lambda\le{\tilde{\sigma}^2(z)}/{4},\mbox{ and }\\
	\mu(z)\;=\;1&\mbox{if}
	&\lambda\le\tilde{\sigma}^2(z)\pi^2(z), \nonumber
\end{eqnarray}
where
\begin{multline} \label{eq: new sigma}
	\tilde{\sigma}^2(Z) = E(X^2|Z) - 2\tilde{Z}^T E(X|Z)  \left[ \int \mu(z) H^{(1)}_{+} \tilde{z}\tilde{z}^T f(z) dz \right]^{-1} 
	\cdot \int \mu(z) H^{(1)}_{+} \tilde{z}  E(X | z) f(z) dz \\
	+  \int \mu(z) H^{(1)}_{+} \tilde{z}^T  E(X |z) f(z) dz  
	\cdot \left[ \int \mu(z) H^{(1)}_{+} \tilde{z}\tilde{z}^T f(z) dz \right]^{-1} \tilde{Z} \tilde{Z}^T \left[ \int \mu(z) H^{(1)}_{+} \tilde{z}\tilde{z}^T f(z) dz \right]^{-1} \\
	\cdot \int \mu(z) H^{(1)}_{+} \tilde{z}  E(X | z) f(z) dz.
\end{multline}
Note $\tilde{\sigma}^2(Z)$ depends on the first and second conditional moments of $X$ given $Z$, which must be known or estimated in previous studies.
The expression of $\tilde{\sigma}^2(Z)$ also depends on the sampling probabilities $\mu(Z)$ on all of the support of $Z$ and thus the optimal $\mu(Z)$ cannot be written explicitly.

Here we propose an algorithm to find the optimal sampling probabilities.
For a fixed value of the penalty parameter $\lambda$, we update iteratively $\tilde{\sigma}^2(Z)$ and $\mu(Z)$ using Algorithm \ref{alg: estimation}, under a maximum number of iterations of $n_{\text{iter}}$ and a convergence criterion $\alpha$.
The term $\pi(Z)$ is estimated by fitting a logistic regression of $Y$ on $Z$ using the data from the first phase of the study.
The integrals in \eqref{eq: new sigma} are estimated with their empirical analogues using observations collected in the first phase of the study.
The first and second conditional moments of $X$ given $Z$ are assumed to be known, at least approximately, from prior studies.
Then we use a binary search to find the value of $\lambda$ such that the constraint
$\int_{\hat{\tilde{\sigma}}^2(z)\ge 4\lambda} \hat{f}_Z(z)dz\left(
1\wedge \hat{\tilde{\sigma}}(z)\hat{\pi}(z)/\sqrt{\lambda}
\right) = N/n$ is satisfied within a small tolerance margin, 
where $\hat{f}_Z(z)$ denotes the empirical density of $Z$,
and solve for the corresponding $\hat{\mu}(Z)$.
Finally, given the estimated optimal $\hat{\mu}(Z)$, by optimizing \eqref{eq: max mu H}, the corresponding estimates of $(\eta_1(Z), \eta_0(Z))$ satisfy
\begin{equation} \label{eq: eta_1}
	\hat{\eta}_1(z) :=
	\begin{cases}
		\hat{\mu}(z)/(2 \hat{\pi}(z)), \quad \text{if } \hat{\mu}(z) \in (0, 2\hat{\pi}(z)], \\ 
		1, \quad \quad\quad \quad \quad \quad \text{if } \hat{\mu}(z) \in (2\hat{\pi}(z),1),\\
	\end{cases}
\end{equation}

\begin{equation} \label{eq: eta_0}
	\hat{\eta}_0(z) :=
	\begin{cases}
		\hat{\mu}(z)/(2(1- \hat{\pi}(z))), \quad\quad\quad \;\; \text{if } \hat{\mu}(z) \in (0, 2\hat{\pi}(z)], \\ 
		(\hat{\mu}(z)-\hat{\pi}(z))/(1-\hat{\pi}(z)), \quad \text{if } \hat{\mu}(z) \in (2\hat{\pi}(z),1].\\
	\end{cases}
\end{equation}

\begin{algorithm}[H]
	\caption{Optimal Sampling probabilities for the PCLvalidate estimator} \label{alg: estimation}
	\begin{algorithmic}[1]
		\Procedure{OptimalMu}{$\{Z_i: i = 1, \dots, n\}, \lambda, \alpha, n_{\text{iter}}$}	
		
		\State Let $k=1$
		\Comment{Initialization}
		
		\State Randomly sample initial values for $\{\hat{\mu}^{(1)}(Z_i)\}$ from Uniform[0,1]
		
		\While {$\Delta \ge \alpha$ and $k \le n_{\text{iter}}$}
		
		\State Estimate $\hat{\tilde{\sigma}}^2(Z_i)$ from \eqref{eq: new sigma} with $\mu(Z_i)=\hat{\mu}^{(k)}(Z_i)$
		
		\State Let 
		\begin{eqnarray*}
			\hat{\mu}^{(k+1)}(Z_i) = 0 & \mbox{if} &{\hat{\tilde{\sigma}}^2(Z_i)}/{4}< \lambda, \\
			\hat{\mu}^{(k+1)}(Z_i) = \hat{\tilde{\sigma}}(Z_i)\hat{\pi}(Z_i)/ \sqrt{\lambda} & \mbox{if} & \hat{\tilde{\sigma}}^2(Z_i)\hat{\pi}^2(Z_i)\le\lambda \le {\hat{\tilde{\sigma}}^2(Z_i)}/{4}, \\
			\hat{\mu}^{(k+1)}(Z_i) = 1 & \mbox{if} & \lambda < \hat{\tilde{\sigma}}^2(Z_i)\hat{\pi}^2(Z_i).			
		\end{eqnarray*}
		
		\State Let $r(Z_i) = (1 - \hat{\pi}(Z_i))/\hat{\pi}(Z_i)$ whenever $\hat{\mu}^{(k+1)}(Z_i) \in (0, 2 \hat{\pi}(Z_i))$ and
		$r(Z_i) = (1- \hat{\pi}(Z_i)) / (\hat{\mu}^{(k+1)}(Z_i) - \hat{\pi}(Z_i))$ whenever $\hat{\mu}^{(k+1)}(Z_i) \in (2 \hat{\pi}(Z_i), 1)$
		
		\State Update $H_+^{(1)}$ with new value of $r(Z_i)$
		
		\State Update $\Delta$ as the difference in 
		\eqref{eq: var of beta} $ - \, \lambda \int_{\hat{\tilde{\sigma}}^2(z)\ge 4\lambda} \hat{f}_Z(z)dz\left(
		1\wedge \hat{\tilde{\sigma}}(z)\hat{\pi}(z)/\sqrt{\lambda}
		\right)$ between the current and previous iteration
		
		\State Update $k = k + 1$		
		
		\EndWhile
		
		\State \textbf{return} $\hat{\mu}(Z_i) = \hat{\mu}^{(k)}(Z_i)$
		\Comment{Output $\{\hat{\mu}(Z_i): i = 1,\dots, n\}$}
		\EndProcedure
	\end{algorithmic}
\end{algorithm}

The development here assumed implicitly that the Lagrangian equation has a solution.  When the distributions of the covariates and confounders are discrete, however, there may be only a zero crossing rather than a zero of the equation for $\lambda$.  
In this case, any allocation of $\mu(z)$ along the
marginal values of $z$
where $\tilde{\sigma}^2(z)\pi^2(z)$ first exceeds the zero crossing point of $\lambda$
that also satisfies the constraint on the expected proportion of subjects selected is optimal.

For the PCLboth estimator, the asymptotic covariance matrix of all regression coefficients are given by
$\Sigma_{\text{PCLboth}}= [E(\delta \mathcal{X}  \mathcal{X}^T H^{(1)}_{+}(X, Z) + (1-\delta) \mathcal{T} \mathcal{T}^T H^{(1)}_{-}(Z) ) ]^{-1}$,
where $\mathcal{T}$ and $H^{(1)}_{-}$ are defined previously. The asymptotic variance of $\hat{\beta}$ is given by the reciprocal of
\begin{multline} \label{eq: var of beta PCLboth}
	\int \big[\mu(z) H^{(1)}_{+} E(X^2 | z) f(z) dz  + (1-\mu(z)) H^{(1)}_{-} \hat{R}(z)^2 \big] f(z)dz \\
	- \int \big[\mu(z) H^{(1)}_{+} \tilde{z}^T  E(X | z) + (1-\mu(z)) H^{(1)}_{-}\tilde{z}^T\hat{R}(z)  \big] f(z) dz \\
	\cdot  \left[ \int  \big[ \mu(z) H^{(1)}_{+} \tilde{z}\tilde{z}^T  + (1-\mu(z)) H^{(1)}_{-} \tilde{z}\tilde{z}^T  \big] f(z) dz \right]^{-1} \\
	\cdot \int \big[\mu(z) H^{(1)}_{+} \tilde{z}  E(X | z) + (1-\mu(z)) H^{(1)}_{-} \tilde{z}\hat{R}(z) \big] f(z) dz.
\end{multline}
Both $H^{(1)}_{+} $ and $H^{(1)}_{-} $ involve the sampling probabilities $\eta_1(Z)$ and $\eta_0(Z)$ and there is no explicit expression of the optimal $\mu(Z)$, $\eta_1(Z)$ and $\eta_0(Z)$. 
In addition, \eqref{eq: var of beta PCLboth} includes the term $\hat{R}(Z)$ which contains the unknown parameter $\beta$.
As a result, it is not feasible to use an algorithm similar to that for the PCLvalidate estimator to estimate the optimal $\mu(Z)$.
Empirical evidence from later sections of the simulation studies demonstrates that the optimal sampling scheme for the PCLvalidate estimator performs well when combined with the PCLboth estimator. Hence, it is recommended to use the same optimal sampling scheme designed for the PCLvalidate estimator when estimating the predictor effect using the PCLboth estimator.

\subsection{A Comparison of Optimal Sampling Schemes}

When $E(X|Z)$ can be represented as a linear function of $Z$, \eqref{eq: new sigma} degenerates to the conditional variance $\sigma^2(Z) = \text{Var}[X|Z]$. 
To see this, suppose $E(X|Z) =  \tilde{Z}  \rho$, where $\tilde{Z}  = (1, Z^T)^T$ and $\rho$ is the vector of linear coefficients.
Then it can be readily shown that
$\left[ \int \mu(z) H^{(1)}_{+} \tilde{z}\tilde{z}^T f(z) dz \right]^{-1} 
\cdot \int \mu(z) H^{(1)}_{+} \tilde{z}  E(X | z) f(z) dz = \rho$, and $\tilde{\sigma}^2(Z) = E(X^2|Z)  - [E(X|Z)]^2 = \sigma^2(Z)$.
Intuitively, the quantity
$\left[ \int \mu(z) H^{(1)}_{+} \tilde{z}\tilde{z}^T f(z) dz \right]^{-1} 
\cdot \int \mu(z) H^{(1)}_{+} \tilde{z}  E(X | z) f(z) dz $ 
is the coefficient of regressing $X$ on $Z$, re-weighted by the measure $\mu(z) H^{(1)}_{+} f(z)dz$.
If $E(X|Z) =  \tilde{Z}  \rho$,
the regression coefficients remain unaffected by the change of measure.

To facilitate comparison between the two sampling schemes, we briefly describe the result in \cite{Tao2020} using the same notation as in this paper. 
By maximizing the expected Fisher information of $\beta$ at the second phase, as per \cite{Robins1994}, subject to the constraint of selecting an expected proportion of subjects, their optimal $\mu(Z)$ satisfies
\begin{eqnarray} \label{eq: soln of mu}
	\mu(z)\;=\;0&\mbox{if}
	&{\sigma^2(z)}/{4}< \lambda,\\ \nonumber
	\mu(z) =\;\sigma(z)\pi(z)/ \sqrt{\lambda}
	&\mbox{if}
	&\sigma^2(z)\pi^2(z)\le\lambda\le{\sigma^2(z)}/{4},\mbox{ and }\\
	\mu(z)\;=\;1&\mbox{if}
	&\lambda\le\sigma^2(z)\pi^2(z), \nonumber
\end{eqnarray}
where $\lambda$ satisfies
$$
\int_{\sigma^2(z)\ge 4\lambda}f_Z(z)dz\left(
1\wedge \sigma(z)\pi(z)/\sqrt{\lambda}
\right) = N/n.
$$ 

%Note the above solutions are based on the assumption, not unreasonable in a case-control setting, that $\pi(z)$ is everywhere less than $1/2$.
%Upon solving the optimal $\mu$, subjects can be selected according to the conditional probabilities $\eta_1(z)$ and $\eta_0(z)$ given by \eqref{eq: eta_1} and \eqref{eq: eta_0}.
Comparing the expressions \eqref{eq: soln of mu2} and \eqref{eq: soln of mu}, it is easy to see that
when $\tilde{\sigma}^2(Z) = \sigma^2(Z)$, the optimal sampling probabilities $\mu(Z)$ in our proposed method unify with the optimal $\mu(Z)$ for efficiency in testing and estimating local effects developed by \cite{Tao2020}. 

If $E(X|Z)$ is not linear with respect to $Z$, however, \eqref{eq: new sigma} is generally not the same as the conditional variance of $X$ given $Z$, and the proposed optimal sampling probabilities $\mu(Z)$ are different from those in \cite{Tao2020}.
Without assuming local effects $\beta = o(1)$, the proposed approach relies on the extra assumption that both $E(X|Z)$ and $E(X^2|Z)$ are known \textit{a priori}, or at least can be well estimated from previous studies, whereas only $\text{Var}(X|Z)$ needs to be known in \cite{Tao2020}.

\section{Hypothesis Testing Methods} \label{sect: test and estimate}

After selecting the subsample to collect the previously unobserved predictor $X$, the natural subsequent questions are how to estimate the effect of $X$ on $Y$ and how to test the statistical significance of the effect. 
In this section, we briefly discuss two methods available for testing the hypothesis $\beta =0$ given the sampling probabilities $\mu(Z)$ and $(\eta_1(Z), \eta_0(Z))$ of the subsample selection schemes adopted by the researcher. 
%and for estimating the regression coefficients .

\subsection{The Score Test}

Based on the efficient score of the model (\cite{Robins1994}), a score test with statistic
\begin{equation} \label{eq: score test}	
	\hat{S}_{\text{eff}} /
	(\hat{I}_{\beta,\beta} - \hat{I}_{\beta,\alpha} \hat{I}^{-1}_{\alpha,\alpha}
		\hat{I}^T_{\beta,\alpha})^{1/2}
\end{equation}
can be adopted to test the hypothesis $\beta = 0$.
Let $p_1(z) = P(Y=1\mid Z=z, \delta = 1)$, then the efficient score in \eqref{eq: score test} can be estimated by
\begin{multline} 
	\hat{S}_{\text{eff}} = \sum_{i=1}^n(Y_i - \hat{\pi}_\alpha(Z_i))
	\hat{E}^{X|Z}\{X_i|Z_i\}  \\
	+ \sum_{i=1}^n\delta_i(Y_i - \hat{p}_1(Z_i))
	(X_i - \hat{E}^{X|Z}\{X|Z_i\})  
	- \hat{I}_{\beta,\alpha} \hat{I}^{-1}_{\alpha,\alpha}
	\sum_{i=1}^n(Y_i-\hat{\pi}_\alpha(Z_i))Z_i.
\end{multline}
Furthermore, estimate the information components with
$\hat{I}_{\beta,\beta} = \sum_i \hat{\pi}(Z_i)(1 -\hat{\pi}(Z_i)) (\hat{E}^{X|Z}\{X_i|Z_i\})^2 + \hat{\mu}(Z_i)\hat{p}_1(Z_i) (1-\hat{p}_1(Z_i)) \hat{\sigma}^2(Z_i)$,
$\hat{I}_{\beta,\alpha} = \sum_i \hat{\pi}(Z_i)(1 -\hat{\pi}(Z_i)) \hat{E}^{X|Z}\{X_i|Z_i\} Z_i^T$ 
and 
$\hat{I}^{-1}_{\alpha,\alpha} = \sum_i \hat{\pi}(Z_i)(1 -\hat{\pi}(Z_i)) (Z_i Z_i^T)$, with $\hat{\pi}_\alpha(Z_i)$, $\hat{p}_1(Z_i)$, and $\hat{E}^{X|Z}\{X_i|Z_i\} $ estimated from the sample. Under the null hypothesis of $\beta = 0$, the test statistic \eqref{eq: score test} follows asymptotically a standard normal distribution.

\subsection{Wald $t$-Tests Derived From the PCL estimators}

Alternatively, Wald-type $t$ statistics derived from the two PCL estimators and their asymptotic covariance matrices (\cite{Wang1997, Wang2002}) can be used to test the null hypothesis. 
Specifically, with the estimated sampling probabilities $\{\hat{\mu}(Z_i)\}$ used in collecting the new predictor,
the Wald $t$-statistics associated with both estimators can be written as 
$\hat{\beta}/\text{se}(\hat{\beta})$,
where $\text{se}(\hat{\beta})$ is the estimate of the reciprocal of \eqref{eq: var of beta} for the PCLvalidate estimate of $\beta$, or of the reciprocal of \eqref{eq: var of beta PCLboth} for the PCLboth estimate of $\beta$.

\section{Simulation Studies} \label{sect: simulation}

\subsection{Simulation Settings}

In this section we conduct a comprehensive simulation study to examine the proposed sampling scheme, and to compare it with existing methods, including the optimal subsample selection for testing local alternatives in \cite{Tao2020}, a purely random selection and a case-control sampling scheme.
With each sampling scheme, we estimate the logistic regression coefficients using four estimators: the naive complete data estimator, the inverse probability weighting (IPW) estimator, and the two pseudo conditional likelihood PCLboth and PCLvalidate estimators described in Section \ref{sect: PCL estimators}.
In addition, we test the null hypothesis $\beta = 0$ using the Wald $t$-tests associated with each of the four estimators as well as the score test introduced in Section \ref{sect: test and estimate}.

% description of settings

We investigate multiple simulation settings to examine the performance of difference sampling schemes and estimators under a variety of scenarios. In each setting, we also examine the results when varying the number of subjects selected in the second phase sub-sample, as well as the event rate $P(Y_i = 1)$ by changing the intercept value in the generative model for the outcome $Y$. 

A total of $50$ independent simulation runs are conducted for each combination of parameters and simulation setting. Each simulation run generates a sample of $n = 400$ subjects from a logistic model. The binary outcome is determined by the predictor of interest $X$ and other covariates $Z$ including the intercept.
The specific parameter values used to generate $X$, $Z$, and $Y$ in each simulation are provided in the respective subsections of the results.

In each simulation run, we conduct the following two-phase sampling procedure.
In the first-phase sample, $X$ is assumed not observable in any subject. 
In the second phase, a sub-sample with expected size $N $ is to be selected to further collect values of $X$.
Assuming the conditional first and second moments of $X$ given $Z$ can be estimated in a pilot study, 
one can calculate the sampling probabilities $\mu(Z_i)$ and $(\eta_1(Z_i), \eta_0(Z_i))$ for each subject $i$ given the observed $\{Z_i\}$ and outcomes $\{Y_i\}$ using the proposed approach described in Section \ref{sect: optimal_PCL}. 
Then the sampling indicators $\{\delta_i\}$ of whether $X_i$ should be collected are generated from independent Bernoulli distributions with probabilities $\eta_1(Z_i)$ or $\eta_0(Z_i)$, depending on whether the observed outcome $Y_i$ equals 1 or 0.
We also calculate the optimal sampling probabilities given in the method of \cite{Tao2020} and generate the corresponding sampling indicators, assuming the conditional variance of $X$ given $Z$ can be estimated.
In addition, a purely random selection and a case-control sampling are conducted, with the same expected sample size as in the proposed sampling scheme, to collect values of $X$.
In the purely random selection, subjects are selected with equal probabilities regardless of the values of the outcome and covariates.
In the case-control selection scheme, an equal number of subjects are randomly selected from the subgroups of subjects with $Y=1$ and $Y=0$.

We compare the performance of all four sampling schemes in the estimation and hypothesis testing of the effect of $X$ on $Y$.
To evaluate the estimation performance, we compare the mean squared errors (MSE) of the estimates using four estimators: the naive complete data estimator, the inverse probability weighting (IPW) estimator (calculated with R package ``\textit{survey}''), and the two pseudo conditional likelihood estimators PCLboth and PCLvalidate described in Section \ref{sect: test and estimate}. 
When using the PCLvalidate estimator, if $\eta_0(Z_i)$ or $\eta_1(Z_i)$ equals zero, a small positive value (0.01) is substituted for the zero to prevent infinite values in $\log({\eta_1(Z_i)}/{\eta_0(Z_i)})$ in the estimating equations.

The PCLboth estimator uses data from both the validation set where the $X$ is observed and the non-validation set of subjects without values of $X$. All other estimators only use the data from the validation set.
Applying the PCLboth estimator requires estimating $R(Z_i)$, for which we fit a linear or logistic regression of $X$ on $Z$, depending on whether $X$ is continuous or dichotomous, with data in the validation set.
The fitted models are correctly specified for the generative model of $X$ given $Z$ in the simulation settings.
In practice, however, the linear and logistic regression models may be mis-specified, and extra bias can be induced in estimating $R(Z_i)$. 
We will further evaluate the influence of potential mis-specification of the conditional distribution of $X$ given $Z$ in the analysis of real data in Section \ref{sect: data analysis}.
To test the null hypothesis of zero effect of $X$ on $Y$ while adjusting for covariates $Z$, we apply the score test and Wald tests described in Section \ref{sect: test and estimate} to the second-phase data of subjects selected with each of the four sampling schemes, and compare the rates of rejecting the null hypothesis with each approach.

The supplementary material contains additional simulation studies that compare the performance of the sampling schemes in estimating the predictor's effect using the SMLE method based on B-splines as described in (\cite{Tao2017}). 
It is observed that implementing this method can be computationally challenging especially with a large number of covariates and/or potential correlations among them. 
For example, one simulation run with the SMLE method typically takes ten times longer than running a simulation without it.
Thus, for demonstrative purposes, we simplified the simulation setting to include only two covariates and evaluated the SMLE method's performance against other estimating approaches discussed earlier. The results demonstrate that the SMLE method led to larger errors than the other methods and, as a result, is not included in the simulation studies presented in the main paper.

\subsection{Simulation Results}

\subsubsection{Simulation Setting 1}

In the first simulation setting, there are six covariates:
$Z_1 \stackrel{iid}{\sim} \text{Bernoulli}(p = 0.3)$,
$Z_2 \stackrel{iid}{\sim} \text{Uniform}(0,1)$,
$Z_3 \stackrel{iid}{\sim} \text{Bernoulli}(p = 0.7)$,
$Z_4 \stackrel{iid}{\sim} N(0, 1)$,
$Z_5 \stackrel{iid}{\sim} \exp(1)$,
$Z_6 \stackrel{iid}{\sim} \Gamma(0.5, 1)$.
The predictor of interest $X$ is generated from a Bernoulli distribution with $P(X=1) = \exp(-1 + \mathbf{Z} \boldsymbol{\beta}_Z)/(1+\exp(-1 + \mathbf{Z} \boldsymbol{\beta}_Z))$, where $\boldsymbol{\beta}_Z = (0.5, 0.5, 1, 2, -3, -2)$.
The outcome $Y$ is generated with a logistic model with regression coefficients 
$\boldsymbol{\beta} = (\beta_0, 0.5, 0.5, 0, 0, 0, 0.5, \beta_X)$, in which $\beta_0$ is the intercept for controlling the event rate $P(Y=1)$, $\beta_X = 2$ is the coefficient for the predictor of interest $X$, and the rest of the values are regression coefficients for $Z_1$ to $Z_6$.

Table \ref{tb: simu1_MSE} compares the estimation errors with each combination of the four sampling schemes and four estimators. The averaged square roots of the MSE are examined under different scenarios of sample size $N=60, 120$ and $200$, and event rate in the outcome $P(Y=1) = 10\%$ and $15\%$. 
When the sample size is as small as $N=60$, the proposed sampling scheme leads to the best estimation performance with any choice of estimator. PCLboth estimator has the smallest MSE, and the PCvalidation and the naive complete data estimator come next. 
It is worth noting, however, that the naive estimator often leads to large bias in the estimation of other coefficients including the intercept.
The sampling scheme in \cite{Tao2020} and case-control sampling lead to larger errors than the proposed approach, and the purely random sampling method is infeasible with very large errors regardless of the choice of estimator adopted. 
As the second-phase sample size $N$ increases, the MSE decreases with all methods, and the gap in performance among the sampling approaches shrinks. The proposed method remains the best approach overall, leading to smallest errors with most of the estimators. One exception is when applying the sampling method in \cite{Tao2020} and the PCLboth estimator, which sometimes leads to small MSE as well.

\begin{table}[H] 
	\small
	\begin{tabular}{@{}lllllllll@{}}
		\toprule
		\textbf{}      & \multicolumn{4}{l}{\textbf{P(Y=1)=10\%}}                           & \multicolumn{4}{l}{\textbf{P(Y=1)=15\%}}                           \\ \midrule
		\textbf{N=60}  & \textbf{Naive} & \textbf{IPW} & \textbf{PCLboth} & \textbf{PCLval} & \textbf{Naive} & \textbf{IPW} & \textbf{PCLboth} & \textbf{PCLval} \\ \midrule
		Proposed       & 0.891          & 1.061        & 0.715            & 0.911           & 0.821          & 0.964        & 0.541            & 0.84            \\
		TestLocal      & 0.999          & 1.227        & 3.234            & 1.042           & 0.912          & 1.02         & >10        & 0.937           \\
		Random         & >10       & >10     & >10        & >10        & >10       & >10     & >10         & >10        \\
		Case-Control   & 1.837          & 2.226        & 4.48             & 1.995           & 1.764          & 2.021        & 5.407            & 2.011           \\ \midrule
		\textbf{N=120} & \textbf{Naive} & \textbf{IPW} & \textbf{PCLboth} & \textbf{PCLval} & \textbf{Naive} & \textbf{IPW} & \textbf{PCLboth} & \textbf{PCLval} \\ \midrule
		Proposed       & 0.552          & 0.605        & 0.541            & 0.556           & 0.465          & 0.52         & 0.443            & 0.491           \\
		TestLocal      & 0.625          & 0.738        & 0.498            & 0.669           & 0.476          & 0.549        & 0.404            & 0.508           \\
		Random         & 1.406          & 1.406        & >10        & 1.444           & 0.913          & 0.913        & 1.426            & 0.913           \\
		Case-Control   & 0.766          & 0.866        & 0.726            & 0.766           & 0.672          & 0.71         & 0.624            & 0.672           \\ \midrule
		\textbf{N=200} & \textbf{Naive} & \textbf{IPW} & \textbf{PCLboth} & \textbf{PCLval} & \textbf{Naive} & \textbf{IPW} & \textbf{PCLboth} & \textbf{PCLval} \\ \midrule
		Proposed       & 0.435          & 0.417        & 0.406            & 0.413           & 0.394          & 0.446        & 0.439            & 0.439           \\
		TestLocal      & 0.482          & 0.524        & 0.398            & 0.521           & 0.436          & 0.473        & 0.419            & 0.455           \\
		Random         & 0.7            & 0.7          & 0.576            & 0.7             & 0.681          & 0.681        & 0.57             & 0.681           \\
		Case-Control   & 0.526          & 0.547        & 0.515            & 0.526           & 0.553          & 0.571        & 0.536            & 0.553           \\ \bottomrule
	\end{tabular}
	\caption{Simulation setting 1: comparison of averaged square roots of mean squared errors (MSE) using different sampling schemes (rows: proposed sampling scheme, method in \cite{Tao2020} (TestLocal), purely random, case-control) and estimation methods (columns: naive complete data estimator, inverse probability weighting (IPW), PCLboth, PCLvalidate). }
	\label{tb: simu1_MSE}
\end{table}

Table \ref{tb: simu1_sig} compares the results of testing the null hypothesis $\beta_X = 0$ on data sampled with each of the four sampling schemes, and with different hypothesis testing methods: the Wald tests associated with each of the four estimators, and the score test.
When $N=60$, the proposed sampling and the method in \cite{Tao2020} both lead to high rates (around 90\%) of rejecting the null hypothesis, and the score test performs better than the Wald tests. The case-control sampling performs worse, with about 70\% rates of rejecting the null using the Wald tests, and around 90\% rejection rate using the score test. The purely random sampling has the worse performance with much lower rate of rejecting the null.
As sample size increases, the overall rates of rejecting the null increases with all sampling and testing methods, but the same ordering of performance persists.

\begin{table}[H]
	\centering
	\scriptsize
	\begin{tabular}{lllllllllll}
		\hline
		\textbf{}      & \multicolumn{5}{l}{\textbf{P(Y=1)=10\%}}                                            & \multicolumn{5}{l}{\textbf{P(Y=1)=15\%}}                                            \\ \hline
		\textbf{N=60}  & \textbf{Naive} & \textbf{IPW} & \textbf{PCLboth} & \textbf{PCLval} & \textbf{Score} & \textbf{Naive} & \textbf{IPW} & \textbf{PCLboth} & \textbf{PCLval} & \textbf{Score} \\ \hline
		Proposed       & 0.81           & 0.81         & 0.87             & 0.84            & 0.91           & 0.89           & 0.89         & 0.94             & 0.91            & 0.96           \\
		TestLocal      & 0.83           & 0.78         & 0.87             & 0.83            & 0.91           & 0.86           & 0.83         & 0.95             & 0.86            & 0.96           \\
		Random         & 0.2            & 0.43         & 0.4              & 0.16            & 0.62           & 0.34           & 0.43         & 0.57             & 0.33            & 0.77           \\
		Case-Control   & 0.71           & 0.71         & 0.71             & 0.71            & 0.84           & 0.67           & 0.67         & 0.67             & 0.67            & 0.91           \\ \hline
		\textbf{N=120} & \textbf{Naive} & \textbf{IPW} & \textbf{PCLboth} & \textbf{PCLval} & \textbf{Score} & \textbf{Naive} & \textbf{IPW} & \textbf{PCLboth} & \textbf{PCLval} & \textbf{Score} \\ \hline
		Proposed       & 0.9            & 0.92         & 0.94             & 0.94            & 0.94           & 0.97           & 0.98         & 0.99             & 0.99            & 1              \\
		TestLocal      & 0.92           & 0.89         & 0.94             & 0.9             & 0.95           & 0.98           & 0.97         & 0.99             & 0.98            & 1              \\
		Random         & 0.56           & 0.55         & 0.76             & 0.53            & 0.76           & 0.72           & 0.71         & 0.83             & 0.73            & 0.89           \\
		Case-Control   & 0.87           & 0.87         & 0.87             & 0.87            & 0.89           & 0.91           & 0.91         & 0.91             & 0.91            & 0.96           \\ \hline
		\textbf{N=200} & \textbf{Naive} & \textbf{IPW} & \textbf{PCLboth} & \textbf{PCLval} & \textbf{Score} & \textbf{Naive} & \textbf{IPW} & \textbf{PCLboth} & \textbf{PCLval} & \textbf{Score} \\ \hline
		Proposed       & 0.98           & 0.98         & 0.98             & 0.98            & 0.98           & 1              & 1            & 1                & 1               & 1              \\
		TestLocal      & 0.98           & 0.95         & 0.98             & 0.95            & 0.98           & 1              & 1            & 1                & 1               & 1              \\
		Random         & 0.8            & 0.8          & 0.87             & 0.79            & 0.88           & 0.94           & 0.94         & 0.97             & 0.94            & 0.96           \\
		Case-Control   & 0.94           & 0.94         & 0.94             & 0.94            & 0.96           & 1              & 1            & 1                & 1               & 1              \\ \hline
	\end{tabular}
	\caption{Simulation setting 1: comparison of rates of rejecting the null hypothesis using different sampling schemes (rows: proposed sampling scheme, method in \cite{Tao2020} (TestLocal), purely random, case-control) and testing methods (columns: Wald tests associated with the naive complete data estimator, inverse probability weighting (IPW), PCLboth, PCLvalidate, and the score test). }
	\label{tb: simu1_sig}
\end{table}

\subsubsection{Simulation Setting 2}

In this simulation setting, we fix $P(Y=1) = 15\%$ and vary the effect size $\beta_X = 0.5$ and $1$ for the predictor of interest on the outcome.
The other model parameters remain the same as in simulation setting 1 in the previous section.
Results of the estimation and hypothesis testing using different sampling schemes under different sample size requirements ($N=80, 120, 200$) are shown in Tables \ref{tb: simu2_MSE} and \ref{tb: simu2_sig}. In this setting, the smallest sample size is increased from 60 to 80 to accommodate for the smaller effect size for approaches to be feasible.
To demonstrate the difference in sampling probabilities, as an example, the estimated optimal $\mu(Z_i)$ in the $\beta_X = 0.5$ scenario using the proposed method and the method described in \cite{Tao2020} are displayed in additional figures in the Supplementary Material.

With a moderate effect size $\beta_X = 1$, at the largest sample size $N=200$, all four sampling schemes lead to meaningful results, except for the purely random sample combined with the PCLboth estimator, in which some simulation runs have very large MSE. The proposed sampling approach generally performs the best, with small MSE when using all four estimators. The sampling scheme proposed in \cite{Tao2020} leads to similar performance when using the PCLboth estimator. The case-control sampling scheme performs worse than the proposed scheme and the method in \cite{Tao2020}, but is overall better than the purely random sampling. Again we observe the choice of sampling scheme plays a more important role than the choice of estimator. As sample size decreases, the purely random sampling scheme becomes less reliable, leading to larger MSE with all four estimators. The MSEs of the case-control sampling also increase considerably. Both the proposed sampling and the method in \cite{Tao2020} are able to hold up. Overall the proposed approach outperforms other sampling methods when using all estimators. When using the PCLboth estimator, the sampling method in \cite{Tao2020} also demonstrates comparable and sometimes even smaller MSE, indicating the robustness of the PCLboth estimator.

With a smaller effect size $\beta_X = 0.5$, the MSE is generally larger than those in the $\beta_X = 1$ scenario. The proposed sampling scheme out-performs other methods, with the smallest MSE using the PCLboth estimator when sample size is small ($N=80$). The sampling method in \cite{Tao2020} lead to larger MSE than the proposed method, with very large errors in some simulation runs when using the PCLboth estimator. The case-control sampling scheme performs worse than the previous two approaches, and the purely random sampling becomes highly unreliable with very large errors in small-sample scenario.

\begin{table}[H]
	\small
	\begin{tabular}{@{}lllllllll@{}}
		\toprule
		\textbf{}      & \multicolumn{4}{l}{\textbf{$\beta_X$ = 1}}                              & \multicolumn{4}{l}{\textbf{$\beta_X$ = 0.5}}                            \\ \midrule
		\textbf{N=80}  & \textbf{Naive} & \textbf{IPW} & \textbf{PCLboth} & \textbf{PCLval} & \textbf{Naive} & \textbf{IPW} & \textbf{PCLboth} & \textbf{PCLval} \\ \midrule
		Proposed       & 0.464          & 0.526        & 0.442            & 0.477           & 0.578          & 0.64         & 0.527            & 0.614           \\
		TestLocal      & 0.531          & 0.609        & 0.386            & 0.565           & 0.749          & 0.822        & >10         & 0.776           \\
		Random         & 4.22       & 4.22     & >10         & 4.42        & >10         & >10       & >10            & >10           \\
		Case-Control   & 0.874          & 0.988        & 0.689            & 0.906           & 0.82           & 0.933        & 0.726            & 0.82            \\ \midrule
		\textbf{N=120} & \textbf{Naive} & \textbf{IPW} & \textbf{PCLboth} & \textbf{PCLval} & \textbf{Naive} & \textbf{IPW} & \textbf{PCLboth} & \textbf{PCLval} \\ \midrule
		Proposed       & 0.493          & 0.513        & 0.465            & 0.489           & 0.621          & 0.61         & 0.579            & 0.607           \\
		TestLocal      & 0.521          & 0.591        & 0.434            & 0.569           & 0.612          & 0.654        & 0.559            & 0.654           \\
		Random         & 1.846          & 1.846        & 3.361            & 2.006           & 3.626          & 3.626        & 19.507           & 4.135           \\
		Case-Control   & 0.644          & 0.689        & 0.625            & 0.644           & 0.755          & 0.797        & 0.736            & 0.755           \\ \midrule
		\textbf{N=200} & \textbf{Naive} & \textbf{IPW} & \textbf{PCLboth} & \textbf{PCLval} & \textbf{Naive} & \textbf{IPW} & \textbf{PCLboth} & \textbf{PCLval} \\ \midrule
		Proposed       & 0.448          & 0.451        & 0.428            & 0.452           & 0.454          & 0.442        & 0.428            & 0.435           \\
		TestLocal      & 0.49           & 0.519        & 0.416            & 0.509           & 0.505          & 0.54         & 0.427            & 0.519           \\
		Random         & 0.646          & 0.646        & >10           & 0.646           & 1.042          & 1.042        & >10         & 1.104           \\
		Case-Control   & 0.53           & 0.554        & 0.519            & 0.53            & 0.568          & 0.589        & 0.561            & 0.568           \\ \bottomrule
	\end{tabular}
	\caption{Simulation setting 2: comparison of averaged square roots of mean squared errors (MSE) using different sampling schemes (rows: proposed sampling scheme, sampling method in \cite{Tao2020} (TestLocal), purely random, case-control) and estimation methods (columns: naive complete data estimator, inverse probability weighting (IPW), PCLboth, PCLvalidate). }
	\label{tb: simu2_MSE}
\end{table}

The results of hypothesis testing are provided in Table \ref{tb: simu2_sig}. Due to the small effect size associated with the predictor of interest, the rates of rejecting the null hypothesis are generally smaller comparing to the previous simulation setting. The proposed method and the sampling scheme in \cite{Tao2020} yield similar performance and have the highest rates of rejection using the score test and the Wald test associated with the PCLboth estimator. Case-control sampling comes next, and the purely random sampling again leads to the worst performance.

\begin{table}[H] 
	\centering
	\scriptsize
	\begin{tabular}{lllllllllll}
		\hline
		\textbf{}      & \multicolumn{5}{l}{\textbf{$\beta_X$ = 1}}                                               & \multicolumn{5}{l}{\textbf{$\beta_X$ = 0.5}}                                             \\ \hline
		\textbf{N=80}  & \textbf{Naive} & \textbf{IPW} & \textbf{PCLboth} & \textbf{PCLval} & \textbf{Score} & \textbf{Naive} & \textbf{IPW} & \textbf{PCLboth} & \textbf{PCLval} & \textbf{Score} \\ \hline
		Proposed       & 0.34           & 0.38         & 0.53             & 0.41            & 0.54           & 0.13           & 0.16         & 0.2              & 0.19            & 0.19           \\
		TestLocal      & 0.33           & 0.32         & 0.48             & 0.35            & 0.5            & 0.18           & 0.15         & 0.2              & 0.21            & 0.18           \\
		Random         & 0.15           & 0.21         & 0.29             & 0.15            & 0.34           & 0.08           & 0.25         & 0.12             & 0.07            & 0.14           \\
		Case-Control   & 0.29           & 0.29         & 0.29             & 0.29            & 0.44           & 0.12           & 0.12         & 0.12             & 0.12            & 0.13           \\ \hline
		\textbf{N=120} & \textbf{Naive} & \textbf{IPW} & \textbf{PCLboth} & \textbf{PCLval} & \textbf{Score} & \textbf{Naive} & \textbf{IPW} & \textbf{PCLboth} & \textbf{PCLval} & \textbf{Score} \\ \hline
		Proposed       & 0.37           & 0.48         & 0.55             & 0.51            & 0.52           & 0.15           & 0.19         & 0.23             & 0.2             & 0.22           \\
		TestLocal      & 0.43           & 0.41         & 0.56             & 0.45            & 0.55           & 0.16           & 0.16         & 0.23             & 0.18            & 0.23           \\
		Random         & 0.21           & 0.24         & 0.31             & 0.22            & 0.33           & 0.1            & 0.18         & 0.13             & 0.09            & 0.15           \\
		Case-Control   & 0.38           & 0.38         & 0.38             & 0.38            & 0.45           & 0.16           & 0.16         & 0.16             & 0.16            & 0.18           \\ \hline
		\textbf{N=200} & \textbf{Naive} & \textbf{IPW} & \textbf{PCLboth} & \textbf{PCLval} & \textbf{Score} & \textbf{Naive} & \textbf{IPW} & \textbf{PCLboth} & \textbf{PCLval} & \textbf{Score} \\ \hline
		Proposed       & 0.45           & 0.54         & 0.55             & 0.55            & 0.55           & 0.06           & 0.16         & 0.16             & 0.16            & 0.17           \\
		TestLocal      & 0.62           & 0.55         & 0.55             & 0.56            & 0.55           & 0.15           & 0.08         & 0.18             & 0.07            & 0.18           \\
		Random         & 0.34           & 0.34         & 0.43             & 0.35            & 0.43           & 0.11           & 0.12         & 0.13             & 0.12            & 0.14           \\
		Case-Control   & 0.53           & 0.53         & 0.53             & 0.53            & 0.56           & 0.15           & 0.15         & 0.15             & 0.15            & 0.18           \\ \hline
	\end{tabular}
	\caption{Simulation setting 2: comparison of rates of rejecting the null hypothesis using different sampling schemes (rows: proposed sampling scheme, sampling method in \cite{Tao2020} (TestLocal), purely random, case-control) and testing methods (columns: Wald tests associated with the naive complete data estimator, inverse probability weighting (IPW), PCLboth, PCLvalidate, and the score test). }
	\label{tb: simu2_sig}
\end{table}

\section{Demonstration with Analysis of Covid-19 Data} \label{sect: data analysis}

\subsection{Data}

We demonstrate the proposed method with data from all 170 patients hospitalized for treatment of coronavirus disease 2019 (COVID-19) at University of California San Diego Health between February 10, 2020 and June 17, 2020. For details of the data and a comprehensive analysis of severe vs. mild disease and of recovery of COVID-19, see \cite{Daniels2020}.

For illustrative purposes, here we focus on the binary outcome of whether a patient ever entered the severe disease status, defined as either admission to the ICU or death. We define the outcome $Y=0$ if a patient ever had severe status, and $Y=1$ if the disease was always mild. 
Out of the $170$ patients, $90$ ever entered severe status while $80$ only had mild symptoms. The primary exposure of interest is the use of statins within the 30 days prior to admission (yes or no). Other potentially important covariates included are use of angiotensin-converting enzyme (ACE) inhibitors and angiotensin II receptor blockers (ARBs). Comorbid conditions including obesity, hypertension, CVD (defined as history of coronary artery disease, stroke and/or transient ischemic attack, peripheral arterial disease, or heart failure), diabetes mellitus, and chronic kidney disease (CKD) are also adjusted for in the analysis.

We examine the effectiveness of the proposed sampling scheme for estimating the effect of use of statins on severe vs. mild COVID-19 diseases.
Suppose the predictor of interest, use of statins, was not included in the original data and is unobserved for all of the $n=170$ subjects. 
We simulate the scenario in which the use of statins later becomes of interest and the researcher would like to collect this variable from $N$ of the $170$ existing patients in the second-phase of the study.

To determine which patients are to be selected for gathering information on statins use, the proposed sampling scheme for the PCLvalidate estimator as well as the sampling scheme in \cite{Tao2020}, the purely random selection and the case-control sampling schemes are applied to the data. 
The conditional first and second moments of the predictor given the other covariates as well as the conditional variance are estimated with logistic regressions using the full data so that the optimal sampling probabilities in the proposed approach and in \cite{Tao2020} can be calculated.
To test the null hypothesis $\beta_{\text{statins}} = 0$, we apply both the score test \eqref{eq: score test} and Wald $t$-tests described in Section \ref{sect: test and estimate} to datasets selected using each of the sampling schemes. 
To estimate the regression coefficients, we apply four approaches including the naive complete data estimator, the IPW estimator, and the two PCL estimators, and compare the averaged mean squared errors (MSE) to evaluate the influence of sampling schemes on the estimation accuracy.
Applying the PCLboth estimator requires estimating $R(Z_i)$, which is achieved by fitting a logistic regression of $X$ on the covariates $Z$.
Unlike in the simulation studies, the generative model of $X$ is unknown, and this logistic regression model may be mis-specified. 
Therefore, analyzing the real data provides an opportunity to assess different sampling schemes when using the PCLboth estimator under potential model mis-specification.

We conduct $50$ independent runs of random selections with each of the four schemes.
In comparing the estimation results, logistic regression coefficients estimated using the full data with use of statins observed in all $170$ patients are taken as the ground truth. In the full data, the use of statins is significantly associated with mild vs. severe COVID-19 disease, with estimated $\beta=1.34$ (odds ratio $3.8$) and $p$-value $0.009$.

\subsection{Results}

We compare the estimation results of the effect of statins use with all combinations of sampling schemes and estimators in two settings of $N=40$ and $N=60$.
The first setting represents a scenario of parsimonious sample size, in which less than $25\%$ of the first-phase sample can be selected in the second phase to collect the new predictor of interest.
The second is a modest sample size setting with about $35\%$ of subjects selected in the second stage.

\begin{figure}[H] 
	\begin{minipage}[b]{0.52\linewidth}
		\centering
		\includegraphics[width=\textwidth]{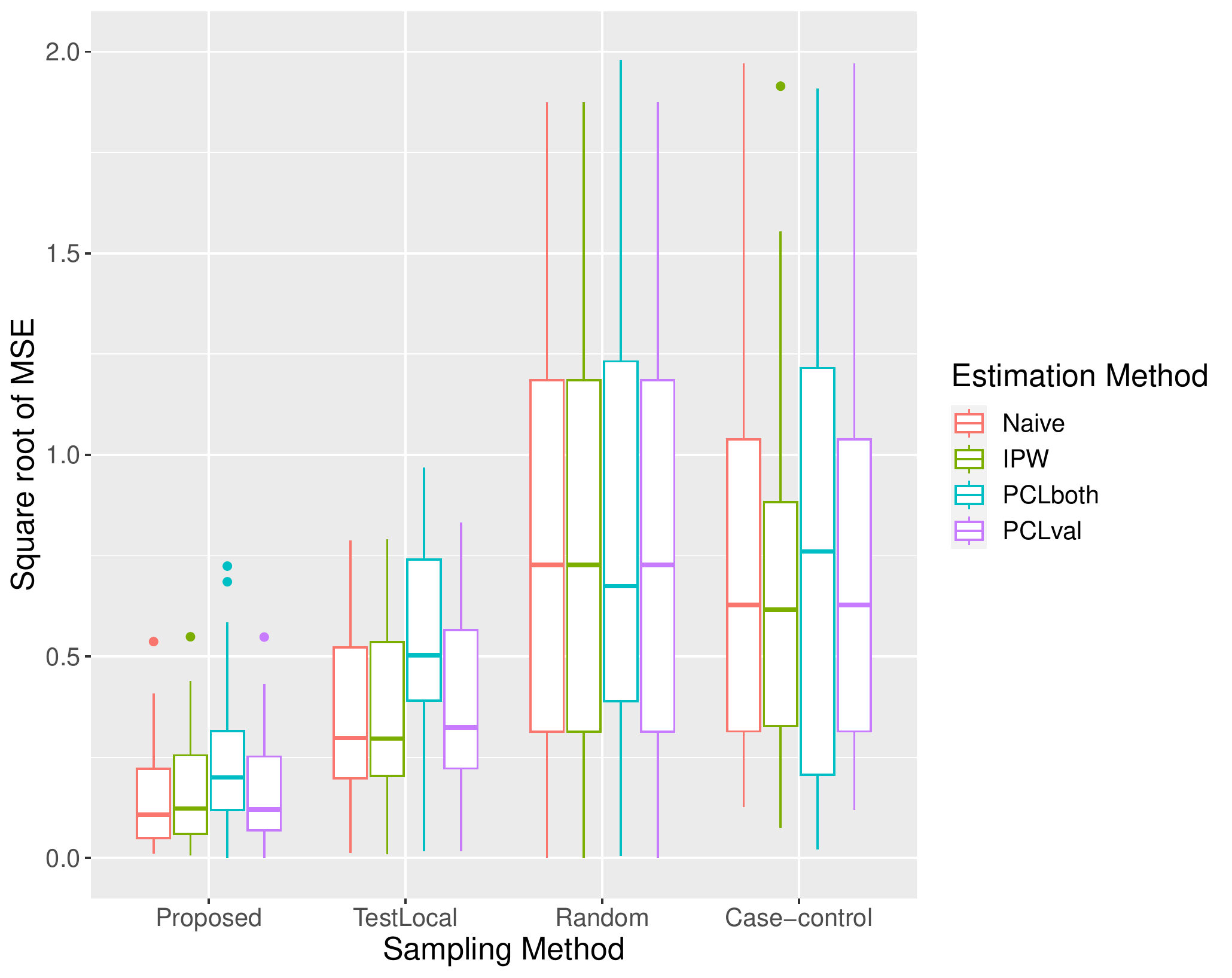}		  
	\end{minipage}
	\hspace{0.5cm}
	\begin{minipage}[b]{0.52\linewidth}
		\centering
		\includegraphics[width=\textwidth]{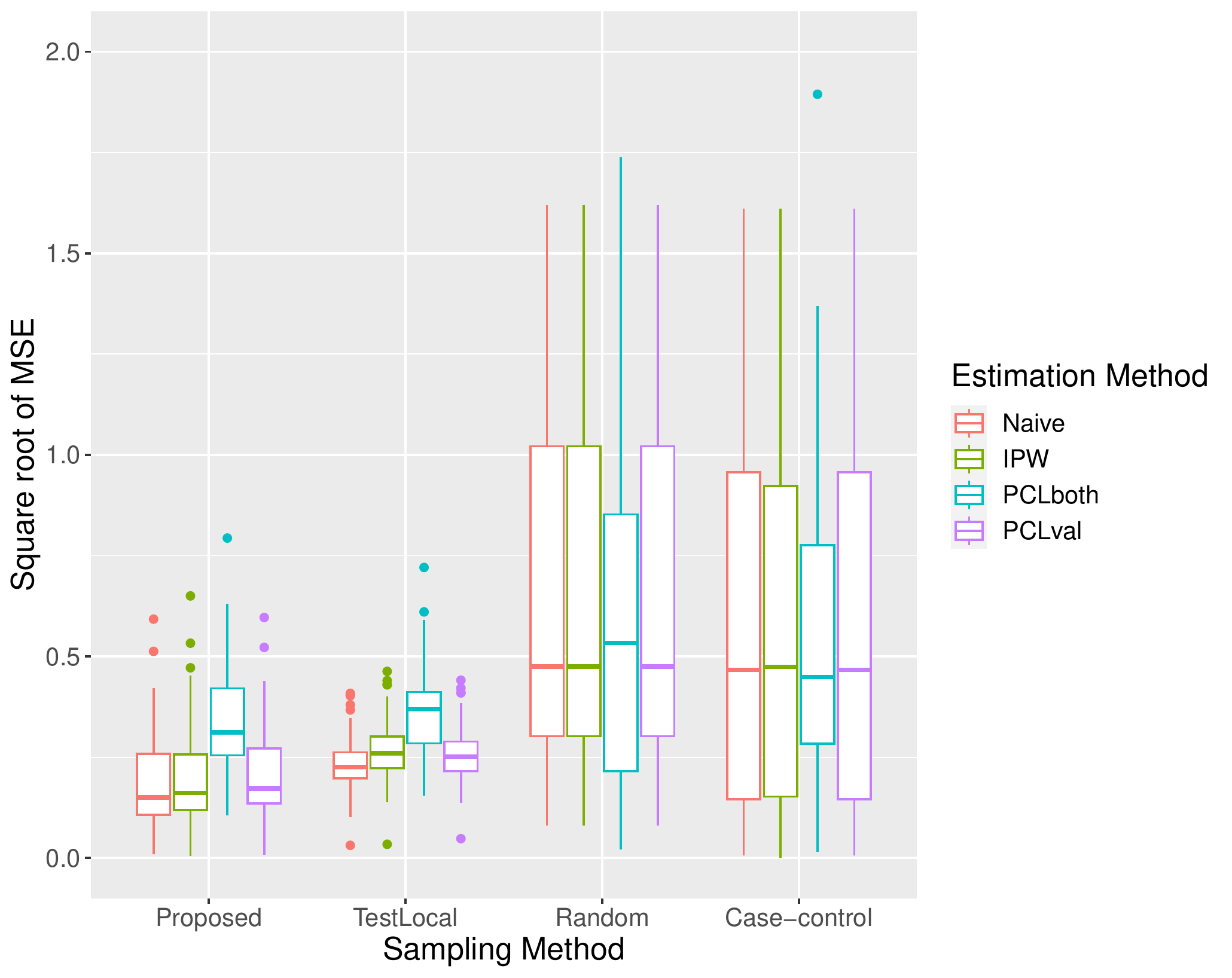}		
	\end{minipage}
	\caption{Comparison of estimation results: $N=40$ and $N=60$} 
	\label{data_est_MSE_plot}
\end{figure}

Figure \ref{data_est_MSE_plot} displays the boxplots of the square roots of MSE in estimating the effect of statins on mild vs. severe diseases in the settings $N=40$ and $N=60$, respectively. 
Table \ref{tab:data_MSE} lists the averaged square roots of MSE over all simulation runs for each combination of sampling schemes and estimators.
In both settings, the proposed sampling scheme leads to an overall better performance with smaller estimation errors with all estimators except for with the PCLboth estimator when $N=60$. 
It is also observed that within each sampling scheme, different estimators yield similar estimation results for the effect of the predictor of interest.
The sampling method in \cite{Tao2020} yields estimation errors about $1.5$ times of those using the proposed sampling scheme when $N=40$, and about $1.3$ times when $N=60$. As the sample size allowance increases, the gap between the to methods decreases.
Both case-control and purely random sampling schemes have much worse performance and larger variability of results. In particular, using the PCLboth estimator can lead to very large errors in some simulation runs, making the method highly unreliable.
Another interesting observation is that when sample size increases, the average errors using the proposed sampling scheme also slightly increase, though still smaller than those of other methods, indicating potential heterogeneity in the effect of the predictor in the cohort of subjects. 
On the other hand, the boxplots demonstrate that the variance of the estimation error generally decreases with increased sample size, as expected.

% add other results to supp material

\begin{table}[H] 
	\centering
	\small
	\begin{tabular}{lllll}
		\hline
		\textbf{N=40} & \textbf{Naive} & \textbf{IPW} & \textbf{PCLboth} & \textbf{PCLval} \\ \hline
		Proposed      & 0.139          & 0.154        & 0.23             & 0.157           \\
		TestLocal     & 0.347          & 0.346        & 0.55             & 0.38            \\
		Random        & 10.755         & 10.755       & 170.38           & 10.755          \\
		Case-Control  & 10.181         & 10.194       & 25.981           & 10.18           \\ \hline
		\textbf{N=60} & \textbf{Naive} & \textbf{IPW} & \textbf{PCLboth} & \textbf{PCLval} \\ \hline
		Proposed      & 0.193          & 0.205        & 0.444            & 0.214           \\
		TestLocal     & 0.239          & 0.271        & 0.36             & 0.261           \\
		Random        & 1.26           & 1.26         & 0.658            & 1.28            \\
		Case-Control  & 2.14           & 2.138        & 129.028          & 2.14            \\ \hline
	\end{tabular} 
	\caption{Comparison of averaged square roots of MSE using each of the sampling schemes and estimators}
	\label{tab:data_MSE}
\end{table}

We also conduct hypothesis tests to assess the effect of statin use on the outcome using Wald t-tests associated with the estimators and the score test. Table \ref{tab:data_tests} displays the rates of rejecting $H_0$ with all combinations of sampling and testing methods. In the $N=40$ setting, the small sample size results in most of the Wald tests having a lower rate of rejecting the null across all sampling schemes. However, even in this setting, the Wald $t$-tests associated with the PCLboth estimator still perform well with rejection rates above $0.95$ when using the proposed sampling scheme and the optimal sampling scheme for testing local alternatives (\cite{Tao2020}). The score test also yields similar performance when paired with the proposed sampling scheme. The purely random and case-control sampling methods largely fail in testing the null hypothesis, indicating they are under-powered when the sample size in the second phase of the study is small.

In the $N=60$ setting, due to the increased sample size in the second phase of the study, both the proposed sampling method and the method in \cite{Tao2020} result in almost 100\% rejection rates of the null hypothesis across all testing methods. However, the purely random and case-control sampling schemes continue to under-perform with maximal rejection rates of around $0.4$ in all testing methods. For both the random and case-control sampling schemes, the score test yields higher rates of rejecting the null compared to other methods.

\begin{table}[H] 
	\centering
	\small
	\begin{tabular}{llllll}
		\hline
		\textbf{N=40} & \textbf{Wald:Naive} & \textbf{Wald:IPW} & \textbf{Wald:PCLboth} & \textbf{Wald:PCLval} & \textbf{Score Test} \\ \hline
		Proposed      & 0.7                 & 0.5               & 0.98                  & 0.84                 & 0.98                \\
		TestLocal     & 0.1                 & 0.2               & 0.98                  & 0.36                 & 0.16                \\
		Random        & 0.08                & 0.36              & 0.08                  & 0.04                 & 0.24                \\
		Case-Control  & 0.04                & 0.38              & 0.12                  & 0.12                 & 0.26                \\ \hline
		\textbf{N=60} & \textbf{Wald:Naive} & \textbf{Wald:IPW} & \textbf{Wald:PCLboth} & \textbf{Wald:PCLval} & \textbf{Score Test} \\ \hline
		Proposed      & 0.98                & 0.92              & 1                     & 1                    & 1                   \\
		TestLocal     & 1                   & 0.96              & 1                     & 0.98                 & 1                   \\
		Random        & 0.3                 & 0.26              & 0.4                   & 0.24                 & 0.46                \\
		Case-Control  & 0.12                & 0.16              & 0.16                  & 0.1                  & 0.2                 \\ \hline
	\end{tabular}
	\caption{Rate of rejecting $H_0$ with each sampling scheme when using the Wald $t$-tests and the score test}
	\label{tab:data_tests}
\end{table}

\section{Discussion}

We have developed here a new approach to optimal selection of subjects for the ascertainment of a new predictor of interest in two-phase designs. The proposed approach aims to select second-phase sample of subjects for estimation of the predictor's effect on the outcome, either local or non-local, using pseudo conditional likelihood estimators.
We derive optimal sampling probabilities for selecting subjects in the second-phase subsample, and compare the results with existing methods including \cite{Tao2020}.

The development here is based on a linear logistic regression model for the effect of the predictor of interest
and the covariates and potential confounders on a dichotomous outcome.  
Central to the optimal sampling probabilities of subjects is the quantity $\tilde{\sigma}^2(Z)$.
Estimating $\tilde{\sigma}^2(Z)$ requires prior knowledge, or at least reasonable estimations of, the quantities
$E(X|Z)$ and $E(X^2|Z)$ from previous studies.
Comparing to the method in \cite{Tao2020} for local effects $\beta = o(1)$ that requires only prior knowledge on $\text{Var}[X|Z]$, 
our proposed approach for estimating general effects requires a small amount of extra knowledge of both the first and second conditional moments of $X$ given $Z$, reflecting the trade-off between the applicability of the approach and assumptions needed to support the applications.
In practice, however, the difficulty of estimating the conditional variance is similar to that in estimating the two first conditional moments, making our proposed approach attractive from a practical point of view given its capacity of dealing with non-local predictor effects.

We assess the effectiveness of our proposed method by conducting simulations and real data analysis, and compare it to existing sampling schemes. Our findings demonstrate that the proposed approach outperforms current methods, particularly when there are stringent sample size requirements for the second-phase subsample and when the model includes a large number of relevant covariates.
Moreover, the data analysis also highlights the performance of various methods under potential model misspecification and complex correlation structures among the covariates and confounders. Results from both the simulations and data analysis reveal the practicality and usefulness of our proposed approach.

If several new predictor variables are to be ascertained, each with a 
different generalized conditional variance $\tilde{\sigma}^2$, a sampling
strategy that minimizes asymptotic variance for one may 
make for less accurate estimation when examining the effects of the others.
Likewise, if transformations of the predictor of interest are perhaps to be chosen as indicated by the data, then which generalized conditional variance to use in computing sampling probabilities is indeterminate. These questions will be addressed in future research.

\bibliographystyle{apalike}
\bibliography{export}

\subsection*{Data availability statement}
The COVID-19 patients data are confidential and are not available to the public.
The simulation data and code that support the findings of this study will be made available in the supplementary material of this article.

\end{document}